\pdfoutput=1

\documentclass{PoS}

\usepackage{cite}

\title{Form factors from lattice QCD}

\ShortTitle{Form factors from lattice QCD}

\author{\speaker{Dru B.\ Renner}\\
Thomas Jefferson National Accelerator Facility (JLab)\\
E-mail: \email{dru@jlab.org}}

\abstract{

Precision computation of hadronic physics with lattice QCD is becoming
feasible. The last decade has seen percent-level calculations of many
simple properties of mesons, and the last few years have seen
calculations of baryon masses, including the nucleon mass, accurate to
a few percent. As computational power increases and algorithms
advance, the precise calculation of a variety of more demanding
hadronic properties will become realistic.  With this in mind, I
discuss the current lattice QCD calculations of generalized parton
distributions with an emphasis on the prospects for well-controlled
calculations for these observables as well. I will do this by way of
several examples:\ the pion and nucleon form factors and moments of
the nucleon parton and generalized-parton distributions.

}

\FullConference{
Sixth International Conference on Quarks and Nuclear Physics\\
April 16-20, 2012\\
Ecole Polytechnique, Palaiseau, Paris}

\begin{document}

\section{Introduction}

Over the last decade, lattice QCD has proven itself capable of
percent-level calculations.  This is a milestone that establishes
lattice field theory as a powerful tool for performing reliable and
quantitative computations of nonperturbative QCD phenomena.  However,
it is important to understand that the computational demands required
for such calculations depend strongly on the quantities of interest.
Consequently, the most impressive lattice calculations to date have
been limited to arguably the simplest QCD observables, mostly meson
properties.  Recently, there has been progress on the determination of
the baryon spectrum.  In particular, the nucleon mass can now be
calculated nonperturbatively from QCD at the few-percent level,
raising the prospects for calculations of more challenging observables
that have long been sought from lattice QCD.

Even with the successful computation of the nucleon mass, the
calculation of nucleon form factors remains a challenge for lattice
QCD.  Significant progress has been made in the last few years.  In
particular, recent computations with nearly physical pion masses
represent a major breakthrough for lattice QCD and have been essential
to the successful determination of the nucleon mass, which is clearly
a necessary first step.  However, there are additional sources of
uncertainty that occur for calculations of nucleon matrix elements.
With nearly physical pion masses becoming more common, the focus is
turning toward fully controlling all uncertainties relevant to the
computation of matrix elements in order to perform calculations of
form factors that can be quantitatively compared to experimental
measurements.

In these proceedings, I start by briefly reviewing several examples of
high-precision lattice calculations and the recent precision
determinations of the masses of baryons, in particular the nucleon.  I
then turn to the pion form factor as a simple example of the more
challenging quantities that our community is interested in.  The
calculation of this observable provides a glimpse of the progress that
we can hopefully expect for nucleon form factors in the years to come.
However, nucleon calculations are still an open issue, so I finish by
looking at just a few representative examples of the generalized
parton distributions of the nucleon.

\section{High-precision lattice QCD calculations}

\begin{figure}
\begin{minipage}{205pt}
\includegraphics[width=205pt]{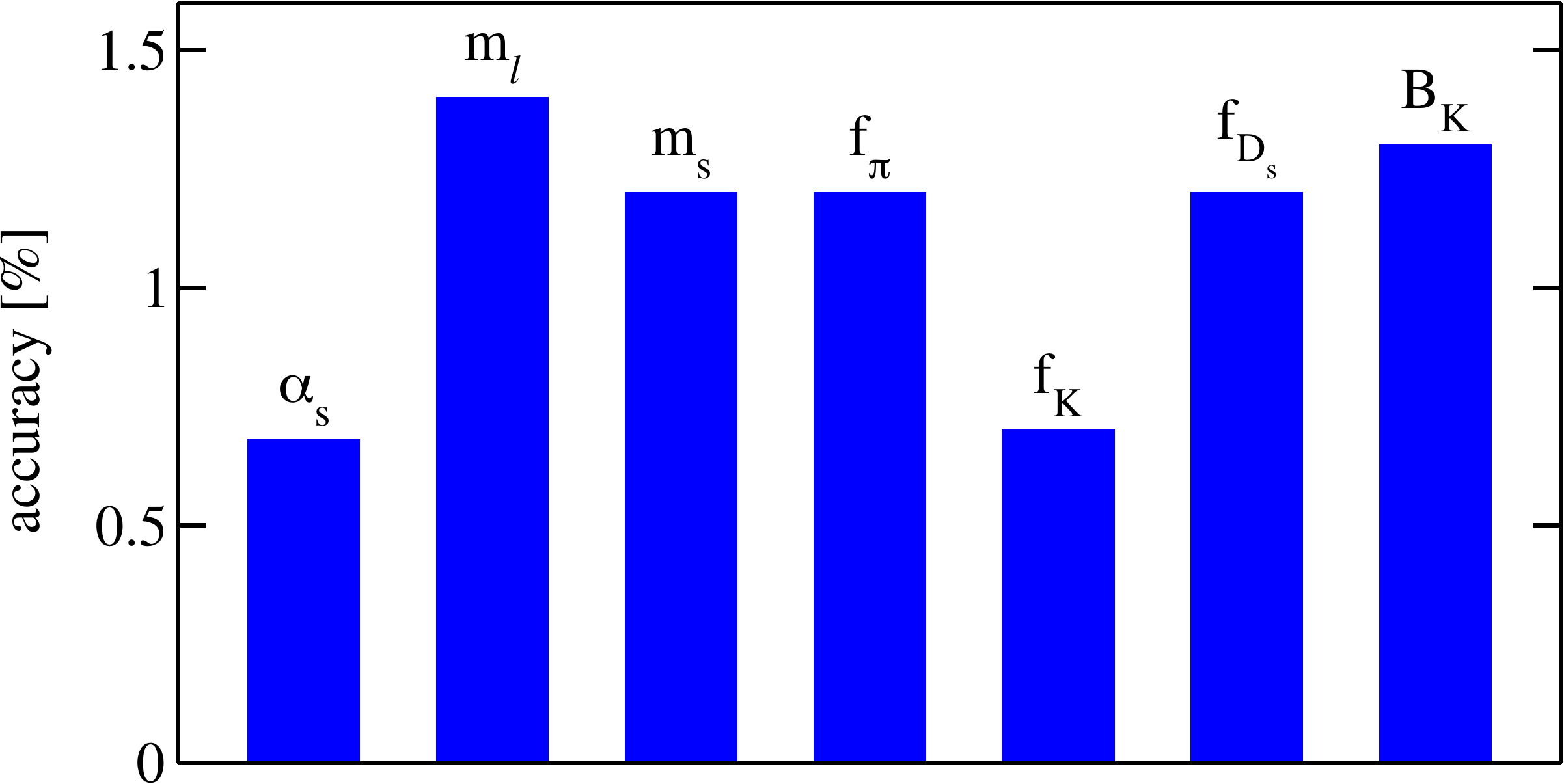}
\caption{Examples of high-precision lattice QCD calculations.  Several
  observables have been calculated with a total precision of one
  percent or better, illustrating that lattice QCD is capable of
  high-precision computations.  The quantities shown here are averages
  of several lattice calculations~\cite{Bethke:2011tr,Laiho:2009eu}.}
\label{hp}
\end{minipage}
\hspace{10pt}
\begin{minipage}{205pt}
\includegraphics[width=205pt]{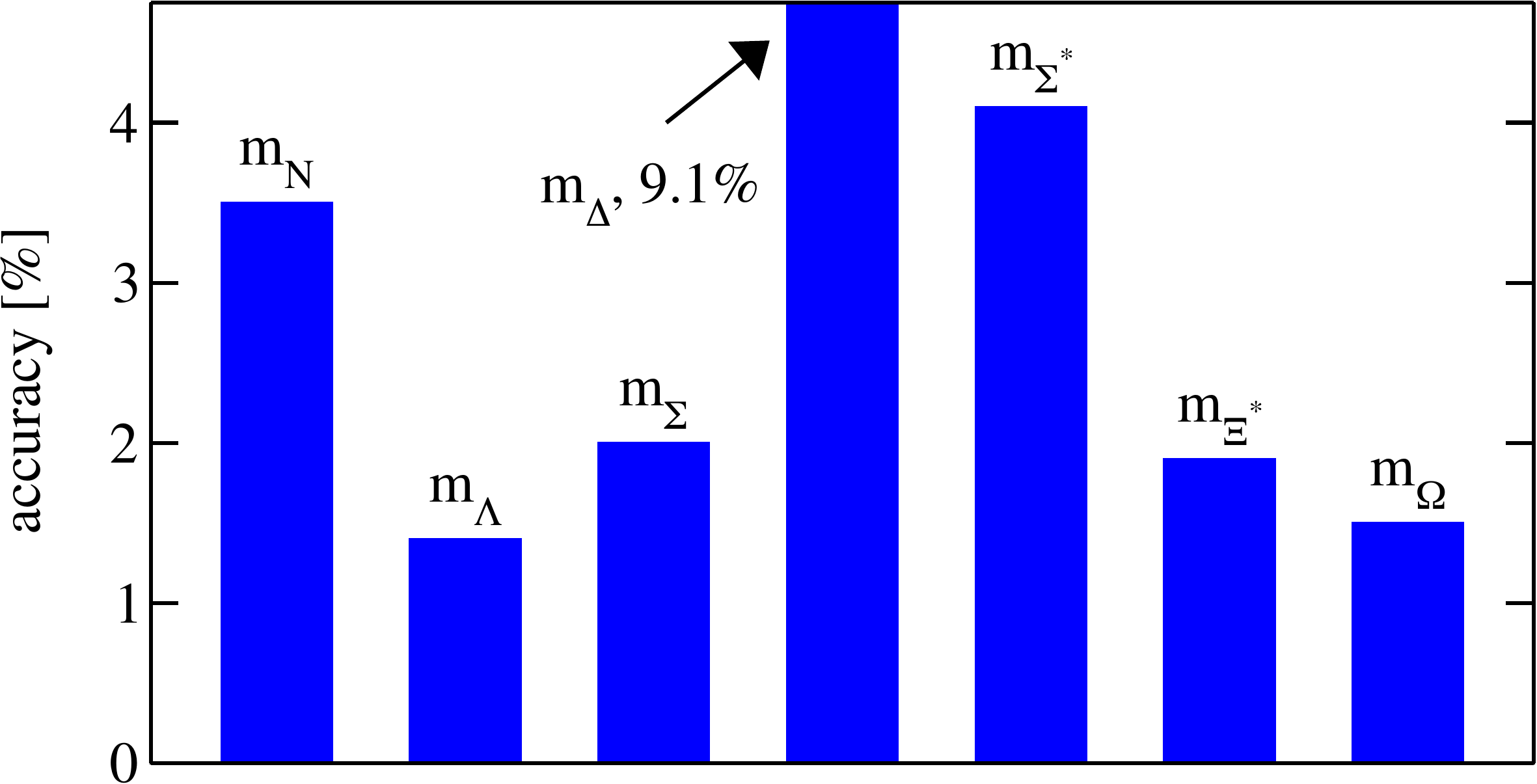}
\caption{An example precision computation of baryon masses.  Lattice
  computations involving baryons are more challenging than mesons and
  have only recently reached the few-percent level.  The masses from
  BMW~\cite{Durr:2008zz} are shown, but PACS-CS~\cite{Aoki:2008sm} and
  ETMC~\cite{Alexandrou:2009qu} find similar results.}
\label{spec}
\end{minipage}
\end{figure}

There is now a small but growing list of quantities that can be
calculated using lattice QCD with percent-level accuracy.
Figure~\ref{hp} shows several.
In order to understand if and when such precision will be brought to
bear on more challenging observables, it is important to note that the
earliest calculations in figure~\ref{hp} were only performed within
the last decade.  Percent-level precision has been reached for
additional quantities since then, but almost all of these are related
to simple properties of stable mesons.  Despite these limitations, the
results in figure~\ref{hp} represent a significant accomplishment of
the lattice community and demonstrate that high-precision calculations
are possible using lattice QCD.

\section{Precision lattice QCD computation of baryon masses}

There is a notable absence of baryon properties in figure~\ref{hp}.
For a variety of reasons, calculations involving baryons require
significantly more computational resources that those for mesons.  In
particular, nucleon properties appear to have a more significant
dependence on the pion masses used for their determinations.  This
represents a computational threshold that afflicts the calculation of
nucleon observables more than others.  Nonetheless, calculations
completed in just the last few years were the first to demonstrate
some quantitative control over the ground-state baryon masses shown in
figure~\ref{spec}.  The determination of the nucleon mass $m_N$ at the
few-percent level is an important advance forward for the lattice
community.  It raises the prospects that the more challenging task of
computing nucleon matrix elements will be within the reach of lattice
QCD.  However, we should bear in mind that the calculation of $m_N$
has only been possible in the last few years.  This suggests that
nucleon matrix elements may very well require yet more time.

\section{Lattice QCD calculation of the pion form factor}

\begin{figure}
\begin{minipage}{205pt}
\includegraphics[width=205pt]{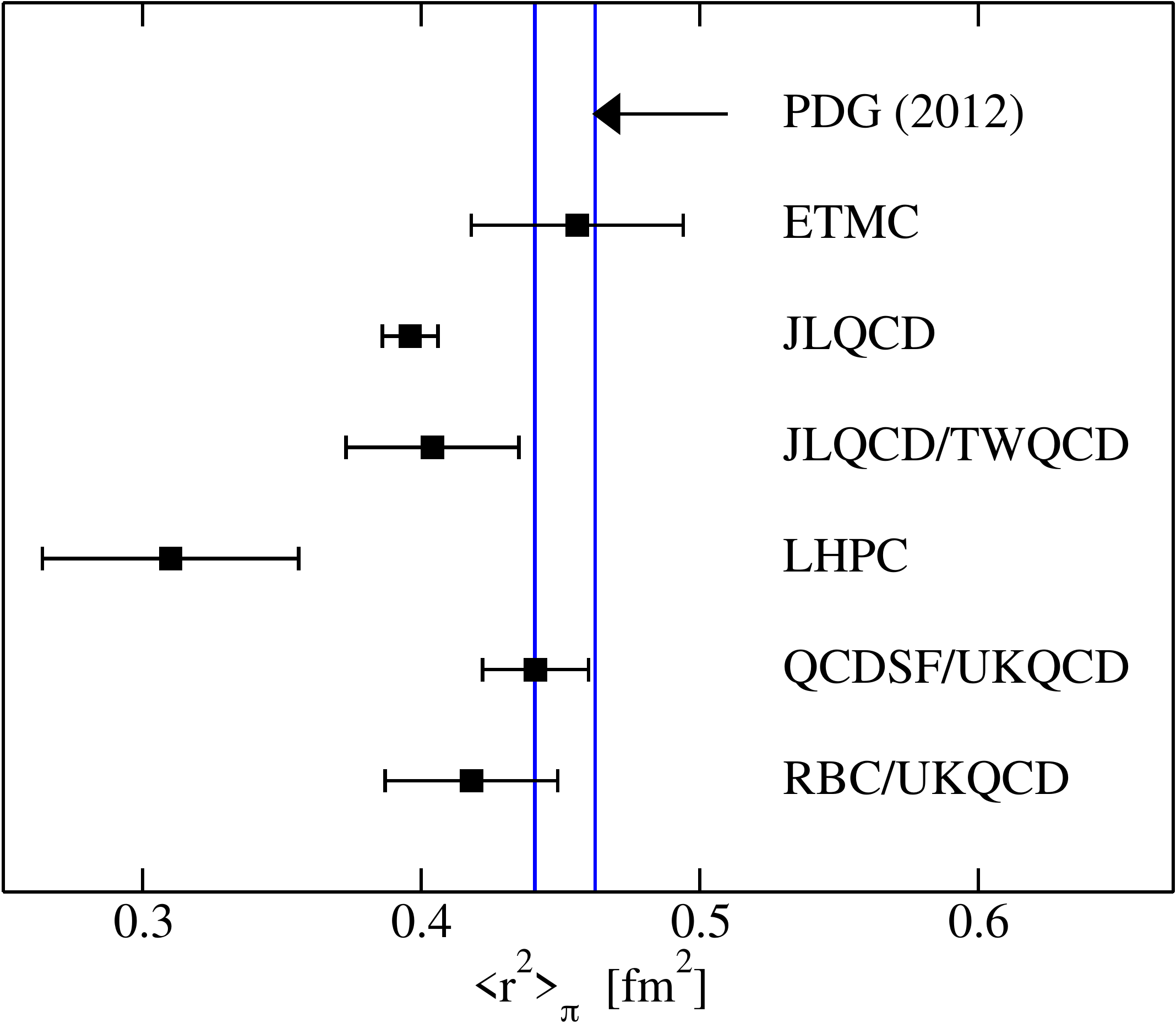}
\caption{An example of a typical but misleading summary plot.  This
  summary of the pion charge radius $\langle r^2\rangle_\pi$, similar
  to the one in~\cite{Frezzotti:2008dr}, illustrates the dangers in
  comparing all lattice computations without regard to controlled
  uncertainties.  The values are from PDG~\cite{pdg:2012},
  ETMC~\cite{Frezzotti:2008dr}, JLQCD~\cite{Hashimoto:2005am},
  JLQCD/TWQCD~\cite{Kaneko:2008kx}, LHPC~\cite{Bonnet:2004fr},
  QCDSF/UKQCD~\cite{Brommel:2006ww} and
  RBC/UKQCD~\cite{Boyle:2008yd}.}
\label{rsq-pion-sum}
\end{minipage}
\hspace{10pt}
\begin{minipage}{205pt}
\includegraphics[width=205pt]{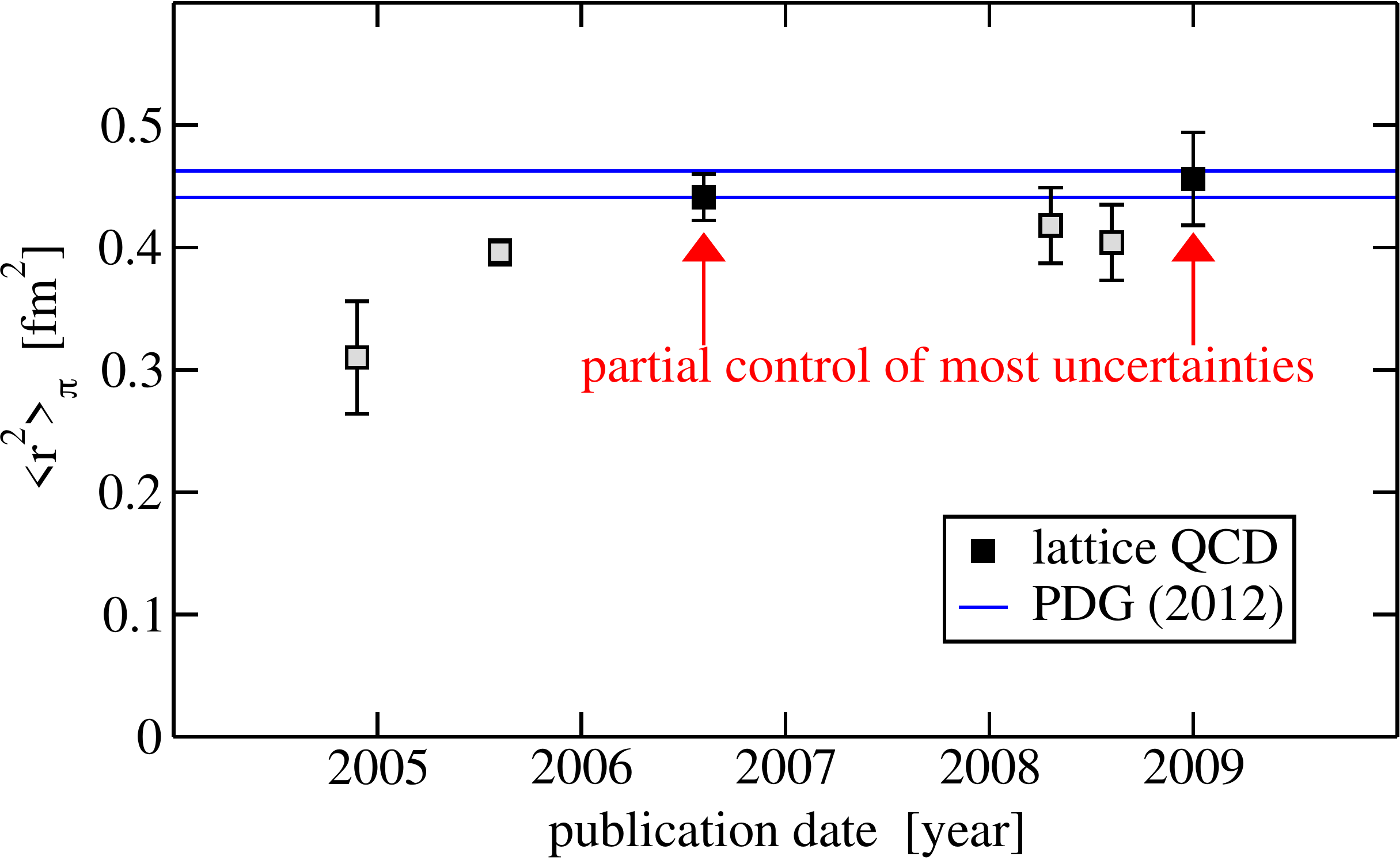}
\caption{Time history of lattice calculations of the pion charge
  radius $\langle r^2\rangle_\pi$.  This time history illustrates that
  lattice QCD calculations with well-controlled uncertainties
  represent definitive predictions of QCD that can be safely compared
  to experimental measurements.  The impression given by this plot
  should be contrasted with that given by
  figure~\protect\ref{rsq-pion-sum}, which fails to distinguish those
  calculations with fully controlled uncertainties.  The values are
  the same as in figure~\protect\ref{rsq-pion-sum} and the two
  highlighted calculations are, starting with the earliest, by
  QCDSF/UKQCD~\cite{Brommel:2006ww} and ETMC~\cite{Frezzotti:2008dr}.}
\label{rsq-th-0}
\end{minipage}
\end{figure}

The pion electromagnetic form factor has recently started to yield to
quantitatively controlled lattice calculations.  It serves as an
example of what to expect for nucleon form factors in the years to
come.  Figure~\ref{rsq-pion-sum} shows a summary plot of lattice QCD
calculations of the charge radius of the pion $\langle
r^2\rangle_\pi$, defined by
\begin{displaymath}
\langle \pi, p | J_\mu | \pi, k \rangle = (p_\mu + k_\mu) F(Q^2)
~~~~~~~~
Q^2=-(p-k)^2
~~~~~~~~
\langle r^2\rangle_\pi = -6 \left. \frac{dF(Q^2)}{dQ^2} \right|_{Q^2=0}\,,
\end{displaymath}
where $J_\mu = \sum_f Q_f J_\mu^f$, $J_\mu^f = \overline{q}_f
\gamma_\mu q_f$ and $Q_f$ is the electric charge for each quark flavor
$f$.  There appears to be no agreement amongst the lattice
calculations, some of which do and some do not agree with the
experimentally measured value of $\langle r^2\rangle_\pi$.  As I will
argue shortly, this plot is misleading, but I have chosen to examine
it as an example of how lattice QCD calculations progress.

To understand what is happening in figure~\ref{rsq-pion-sum}, we can
crudely plot the results for $\langle r^2\rangle_\pi$ versus their
publication date.  This is done in figure~\ref{rsq-th-0}, which shows
a time dependence in the calculations with later results agreeing with
the measured $\langle r^2\rangle_\pi$.  More importantly, after
selecting those calculations that have accounted in some way for the
relevant sources of uncertainty (highlighted in
figure~\ref{rsq-th-0}), we see a consistent picture emerge.  All
well-controlled lattice calculations agree with each other, as should
always happen, and, in this case, agree with the experimental
measurement.  At this point we are left with only two calculations
that meet this burden~\cite{Brommel:2006ww,Frezzotti:2008dr}, but
several of the calculations in figure~\ref{rsq-th-0} are ongoing and
there are new calculations~\cite{Brandt:2011jk} underway.  We should
expect in the next few years to have multiple precise and reliable
lattice calculations of $\langle r^2\rangle_\pi$.  As with the results
in figures~\ref{hp} and \ref{spec}, the easiest observables are
calculated first, but this is to be expected from large-scale
numerical calculations that face multiple computational thresholds
that depend on the quantities of interest.  The calculation of
$\langle r^2\rangle_\pi$ is not only interesting in its own right but
also is a good example of how nucleon form factor calculations may
progress as the computational barriers are crossed for those
observables as well.

\begin{figure}
\begin{minipage}{205pt}
\includegraphics[width=205pt]{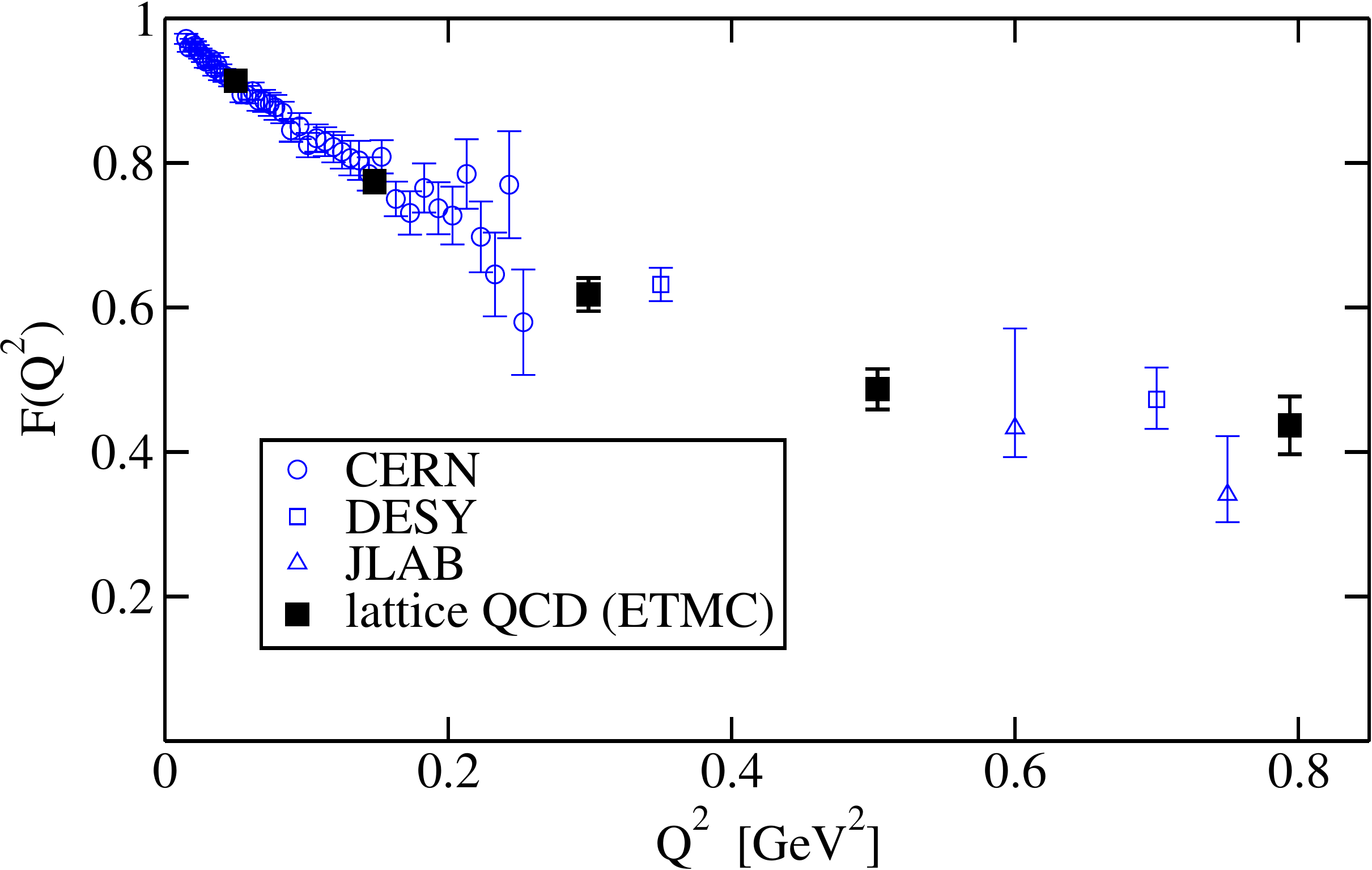}
\caption{Comparison of the pion form factor $F(Q^2)$ from experiment
  and lattice QCD.  ETMC~\cite{Frezzotti:2008dr} has performed a
  lattice calculation with reasonably controlled uncertainties of the
  low $Q^2$ dependence of $F(Q^2)$.  This is an example of the level
  of control sought for the more challenging nucleon form factors.
  The experimental measurements are from CERN~\cite{Amendolia:1986wj},
  DESY~\cite{Ackermann:1977rp,Brauel:1979zk} (reanalyzed
  in~\cite{Tadevosyan:2007yd,Huber:2008id}) and
  JLAB~\cite{Tadevosyan:2007yd}.}
\label{pion-ff}
\end{minipage}
\hspace{10pt}
\begin{minipage}{205pt}
\includegraphics[width=205pt]{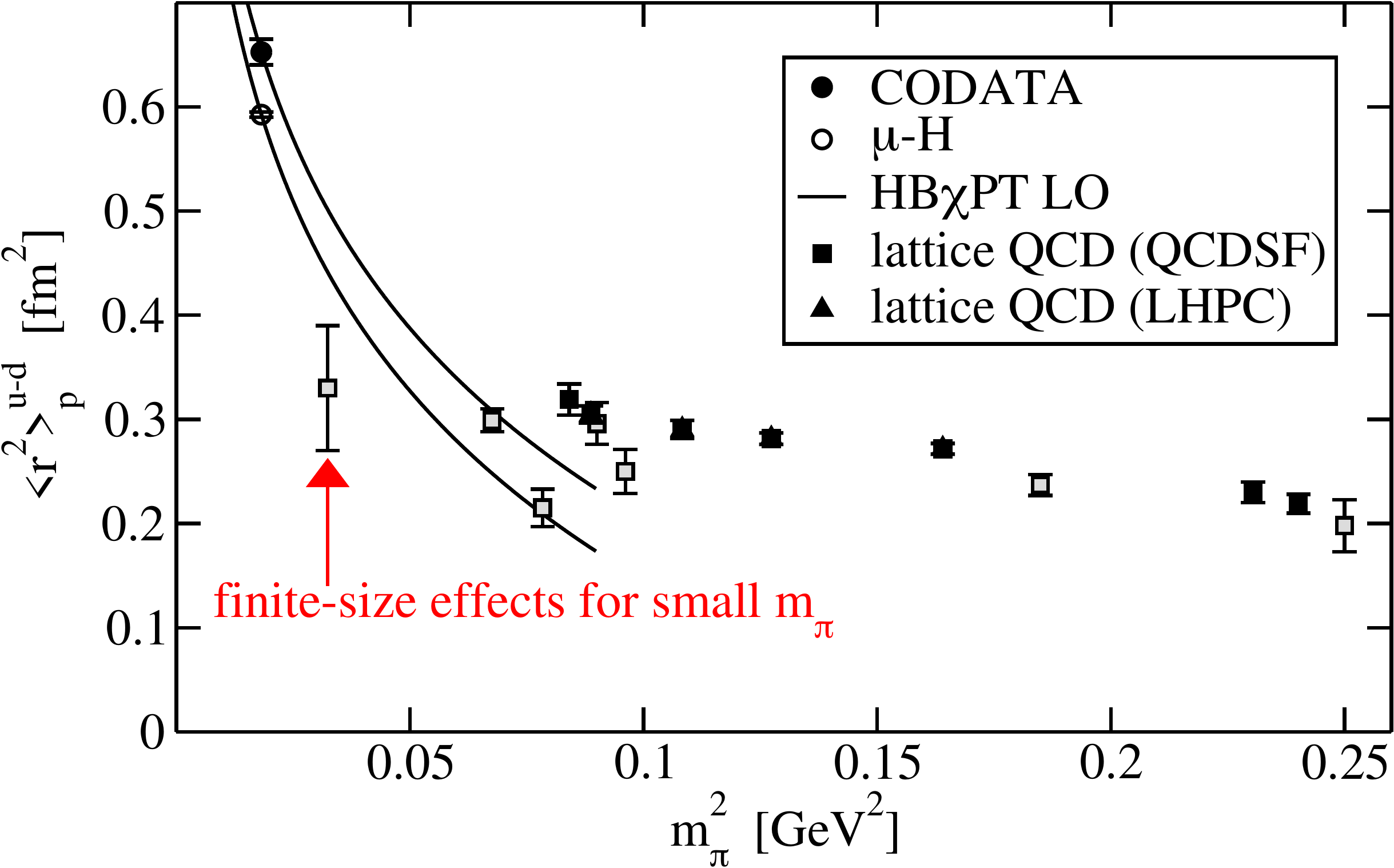}
\caption{The proton isovector charge radius $\langle
  r^2\rangle^{u-d}_p$.  The measured values use CODATA~\cite{codata}
  or $\mu$-$H$~\cite{Pohl:2010zza} for $\langle r^2 \rangle_P$ with
  PDG~\cite{pdg:2012} for $\langle r^2 \rangle_N$ to form $\langle
  r^2\rangle^{u-d}_p = \langle r^2 \rangle_P - \langle r^2 \rangle_N$.
  These are then matched to chiral perturbation
  theory~\cite{Bernard:1992qa}.  The lattice results are from
  QCDSF~\cite{Collins:2011mk} and LHPC~\cite{Syritsyn:2009mx}.
  Requiring well-controlled uncertainties likely eliminates the
  apparent tension between lattice QCD, chiral perturbation
  theory~\cite{Bernard:1992qa} and the measurements.}
\label{rsq-n-3}
\end{minipage}
\end{figure}

Moving beyond the extreme $Q^2\rightarrow 0$ limit characterized by
the slope of the form factor, lattice calculations are also exploring
the non-zero but still low $Q^2$ regime of $Q^2 < 1~\mathrm{GeV}^2$.
Various uncertainties become more challenging with increasing $Q^2$,
but the low $Q^2$ behavior of $F(Q^2)$ can also be calculated with
relatively well-controlled uncertainties.  This is shown in
figure~\ref{pion-ff} along with the experimental results in the low
$Q^2$ region.  The agreement between lattice QCD and experiment is
again rather compelling and bodes well for the eventual determination
of nucleon form factors.

\section{Status of nucleon form factors from lattice QCD}

Lattice calculations with nucleons remain a challenge due to the extra
computational demands that are required.  As such, there is not yet a
compelling success story for lattice determinations of nucleon
structure, but this needs to be understood in context.  It has only
been in the last few years that the lattice community has developed
sufficient algorithms and garnered enough computing power to calculate
even $m_N$ at the few-percent level.  Thus it should not be surprising
that current calculations of nucleon structure have produced mixed
results, with some encouraging agreements and some noticeable
disagreements.  As illustrated by the pion form factor, this is how we
expect lattice calculations to look before fully controlled
uncertainties are achieved.

With this in mind, I simply summarize the current status of nucleon
calculations with a few examples.  To compare with the discussion of
the pion charge radius, I start with the proton charge radius defined
similarly,
\begin{displaymath}
\langle P, p | J^f_\mu | P, k \rangle = K_\mu^1 F^f_1(Q^2) + K_\mu^2 F^f_2(Q^2)
~~~~~~~~
\langle r^2\rangle_p^f = -6 \left. \frac{dF^f_1(Q^2)}{dQ^2} \right|_{Q^2=0}\,,
\end{displaymath}
where $K^{i}_\mu$ are kinematic functions of $p$ and $k$.  From the
outset, I have to make a sacrifice and instead focus on the isovector
radius $\langle r^2\rangle_p^{u-d} = \langle r^2\rangle_p^{u} -
\langle r^2\rangle_p^{d}$.  This restriction eliminates so-called
disconnected diagrams that are an additional source of uncertainty.
Furthermore, I avoid any statements about extrapolations to the
physical point and simply examine the lattice results as functions of
the artificially heavy pion mass $m_\pi$ used in the calculations.
These are shown in figure~\ref{rsq-n-3}.  To avoid unnecessary
clutter, only two lattice calculations are shown as examples, and only
the uncertainties due to numerically integrating the path integral are
shown.  Considering all results in figure~\ref{rsq-n-3} without regard
to the remaining uncertainties in the problem, there would be an
apparent disagreement between lattice calculations and the measured
value.  Additionally, there would seem to be a significant
disagreement with the expectations of heavy baryon chiral perturbation
theory.  However, as with the early calculations for the pion form
factor, the failure to accurately control all uncertainties in the
lattice calculation may very well be the explanation for both of these
observations.

A detailed discussion of the uncertainties for $\langle
r^2\rangle_p^{u-d}$ is beyond the scope of these proceedings, but I
will examine one uncertainty that may be a concern:\ finite-size
effects.  Lattice calculations are performed at a finite physical
volume $L^3$ and correspondingly the allowed momentum modes are
discretized.  Forming the derivative of the form factors requires
taking finite differences in $Q^2$.  Furthermore, this quantity is
expected to diverge in the chiral limit, so we may reasonably expect
an enhanced sensitivity to the volume, which acts as the infrared
regulator for lattice QCD.  Due to the demanding nature of nucleon
calculations, the large volume limit is difficult to study, especially
as $m_\pi$ is decreased.  Consequently, there is not much evidence for
or against a strong finite-size effect for $\langle
r^2\rangle_p^{u-d}$.  In lieu of a full study of the $L$ dependence,
we can require the usual rule of thumb $m_\pi L > 4$.  This
restriction in figure~\ref{rsq-n-3} clearly alleviates the
disagreement with the measured value for $\langle r^2\rangle_p^{u-d}$
and substantially weakens the tension with the expectations from
chiral perturbation theory, suggesting that calculations with larger
$L$ may help resolve these puzzles.

\begin{figure}
\begin{minipage}{205pt}
\includegraphics[width=205pt]{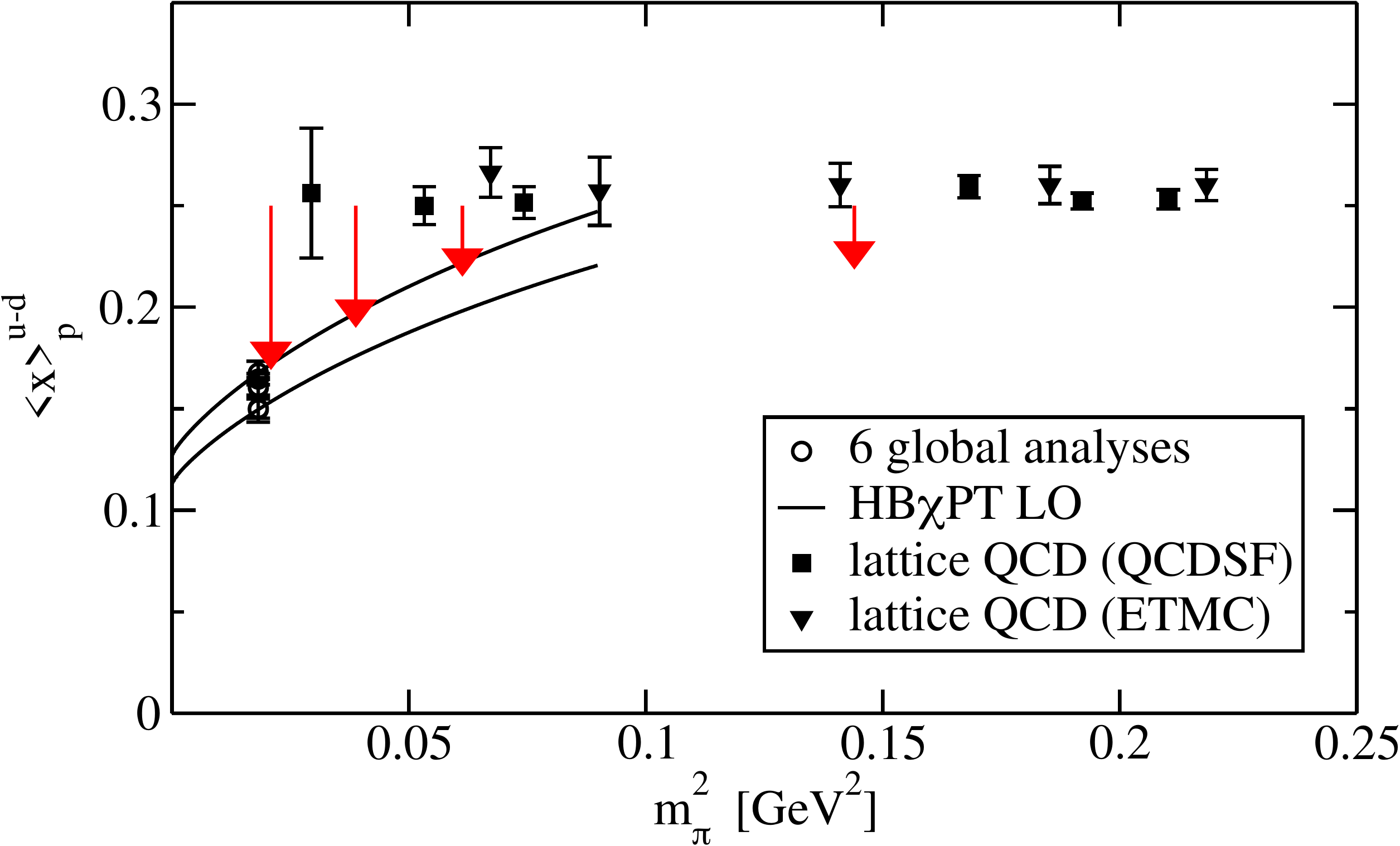}
\caption{The isovector momentum fraction $\langle x \rangle^{u-d}_p$.
  The lattice results are QCDSF~\cite{qcdsf:priv} and
  ETMC~\cite{Alexandrou:2011nr}.  The results from global analyses are
  from~\cite{Renner:2010ks} using~\cite{Alekhin:2009ni, Blumlein:2006be, Blumlein:2004ip,
    JimenezDelgado:2008hf, Martin:2009bu, Alekhin:2006zm}.  The leading-order results from
  chiral perturbation theory~\cite{Arndt:2001ye,Chen:2001eg} are
  matched to the largest and smallest results from the global analyses
  to indicate the impact of the variations of these results.  The
  red arrows indicate the possible impact of the uncertainty,
  suggested by ETMC~\cite{Dinter:2011sg} and LHPC~\cite{Green:2011fg},
  due to excited-state pollution in the lattice calculations as
  described in the text.}
\label{x-1}
\end{minipage}
\hspace{10pt}
\begin{minipage}{205pt}
\includegraphics[width=205pt]{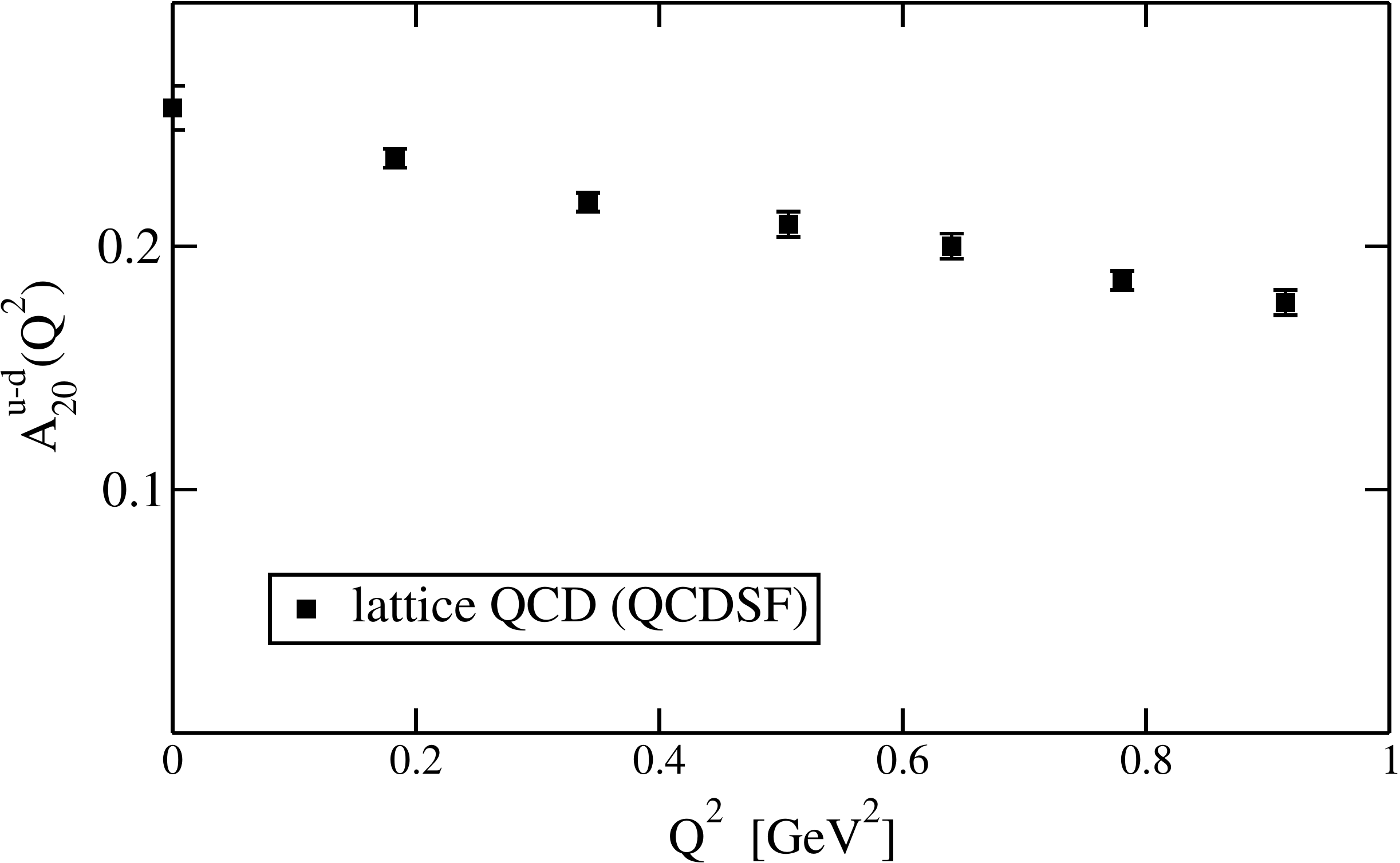}
\caption{The isovector $A^{u-d}_{20}(Q^2)$ generalized form factor.
  The moments of the generalized-parton distributions of the proton
  can be calculated using lattice QCD techniques.  The generalized
  form factor $A_{20}(Q^2)$ shown here is related to the first
  $x$-moment of the sum of the $H(x,\xi,\Delta^2)$ and
  $E(x,\xi,\Delta^2)$ GPDs in which $Q^2=\Delta^2$.  The results
  from~\cite{Sternbeck:2012rw} are shown as an example.  The absence
  of experimental results emphasizes the prospects for genuine
  predictions from lattice QCD once all uncertainties are reliably
  controlled.}
\label{proton-a20}
\end{minipage}
\end{figure}

Nucleon form factors are challenging enough;\ turning to moments of
parton and generalized-parton distributions increases the difficulties
further.  In figure~\ref{x-1},
I show two example calculations of the average longitudinal-momentum
fraction $x$ in the proton $\langle x \rangle^{u-d}_p$ given by
\begin{displaymath}
\langle P, p | O^f_{\mu\nu}| P, k \rangle = 
K^A_{\mu\nu}\, A^f_{20}(Q^2) + K^B_{\mu\nu}\, B^f_{20}(Q^2) + K^C_{\mu\nu}\, C^f_2(Q^2)
~~~~~~~~
\langle x \rangle_P^f = \int_{-1}^1\!\!\!\! dx\,\,\, x\, q^f(x) = A^f_{20}(0)\,,
\end{displaymath}
where the $K_{\mu\nu}^i$ are kinematic functions, $O^f_{\mu\nu}=
\overline{q}_f \gamma_{\left\{\mu\right.} D_{\left.\nu\right\}} q_f$
are symmetric traceless twist-two operators, and $q^f(x)$ can be
related to the quark (anti-quark) parton distribution functions (PDFs)
for $x>0$ ($x<0$).  Generalizations to higher moments $\langle x^n
\rangle^f_P$ are possible.  The isovector combination is used for the
same reasons as discussed regarding $\langle r^2\rangle_p^{u-d}$.
There is an apparent disagreement with the measured value, but again
the issue boils down to fully controlling the relevant uncertainties.
An important uncertainty for $\langle x \rangle^{u-d}_p$ appears to be
a finite-size effect due to excited-state pollution.  (This is a
technical issue related to having a large enough separation between
correlation functions in the calculation.)  In the case of $\langle x
\rangle^{u-d}_p$ there are indications from lattice calculations that
this effect is occurring.  The red arrows in figure~\ref{x-1} show the
sort of corrections suggested by direct calculations.  These
corrections are not universal and can not readily be applied to the
results of other computations, so we can only note that the sign of
the correction and its the $m_\pi$ dependence seems to reduce the
disagreement with the measured $\langle x \rangle^{u-d}_p$.

I have focused on $\langle r^2\rangle_p^{u-d}$ and $\langle x
\rangle^{u-d}_p$ because they are commonly used as benchmarks for
lattice calculations of nucleon structure.  Just as the calculation of
$m_N$ was an essential achievement for making progress, the successful
determination of $\langle r^2\rangle_p^{u-d}$, $\langle x
\rangle^{u-d}_p$ and other similar observables is a necessary step
towards the broader structure program envisioned by the lattice
community.  As just one example of that effort, I consider the
generalized-parton distributions (GPDs).  As for the PDFs, lattice
calculations focus on moments in $x$.  The first moments of the $H$
and $E$ GPDs are given by the generalized form factors $A_{20}(Q^2)$,
$B_{20}(Q^2)$ and $C_{2}(Q^2)$.  One example is
\begin{displaymath}
\int_{-1}^1\!\!\!\! dx\,\,\, x\, \left(H^f(x,\xi,\Delta^2)+E^f(x,\xi,\Delta^2)\right) = A^f_{20}(\Delta^2)+B^f_{20}(\Delta^2)\,,
\end{displaymath}
where the definitions of the $H$ and $E$ GPDs and further details are
available in~\cite{Ji:1996ek}.  There have been several lattice
calculations of these form factors, and one recent example is shown in
figure~\ref{proton-a20}.  The calculation of $A^{u-d}_{20}(Q^2)$ is an
extension of $\langle x \rangle^{u-d}_p$ and all issues relevant for
$\langle x \rangle^{u-d}_p$ are expected to occur for
$A^{u-d}_{20}(Q^2)$ as well.  In fact, $A^{u-d}_{20}(0)=\langle x
\rangle^{u-d}_p$.  One goal of the long-term lattice effort on nucleon
structure is the determination of the low moments of all the nucleon
GPDs.  The restriction to moments in $x$ is a significant limitation
to PDFs, but for GPDs the additional information on the $Q^2=\Delta^2$
dependence adds valuable information on the GPDs that is complimentary
to that which is accessible from experimental measurements.  Thus
well-controlled calculations of the $Q^2$ dependence of form factors
like $A_{20}(Q^2)$ will yield genuine predictions from lattice QCD for
the nucleon and will open several avenues to rich physics topics
including the spin decomposition and transverse structure of the
nucleon.

\section{Conclusions}

In the last decade, lattice QCD has shown itself to be capable of
precision calculations.  The initial successes were understandably
limited to those observables for which all uncertainties could be
controlled with the least computational resources.  However, recent
calculations, such as the determination of $m_N$, indicate that
lattice computations of more demanding hadronic quantities should
become feasible in the years to come.

The pion form factor is currently a well-determined quantity from
lattice QCD, with further improvements expected.  It stands as an
example of the way forward for equally well-controlled lattice
calculations of nucleon structure.  There will be continued emphasis
on benchmark observables, such as $\langle r^2 \rangle^{u-d}_p$ and
$\langle x \rangle^{u-d}_p$, in order to establish control of all
relevant uncertainties for calculations of nucleon matrix elements.
This renewed focus on carefully accessing each uncertainty is already
shedding some light on current puzzles facing lattice calculations of
nucleon structure and will eventually lead to precise and reliable
calculations that can safely be compared to experimental measurements.
However, the real promise of lattice calculations of nucleon matrix
elements ultimately lies in determining observables that lie beyond
the reach of measurements yet offer the possibility of insight into
the deeper mechanisms behind nucleon structure.

\section{Acknowledgements}

This manuscript has been coauthored by Jefferson Science Associates,
LLC under U.S.\ DOE Contract No.\ DE-AC05-06OR23177. The
U.S.\ Government retains a non-exclusive, paid-up, irrevocable,
world-wide license to publish or reproduce this manuscript for
U.S.\ Government purposes.

\bibliographystyle{h-physrev}
\bibliography{qnp2012-renner.bib}

\end{document}